\documentclass[final,5p,times,twocolumn]{elsarticle}

\usepackage{graphicx}
\usepackage{dcolumn}
\usepackage{bm}
\usepackage{color}
\usepackage{bigstrut}
\usepackage{tabularx}
\usepackage{float}
\usepackage{ulem}
\usepackage[]{color}
\usepackage[colorlinks=true]{hyperref}
\usepackage{amsmath}
\newcolumntype{L}[1]{>{\raggedright\arraybackslash}p{#1}}
\newcolumntype{C}[1]{>{\centering\arraybackslash}p{#1}}
\newcolumntype{R}[1]{>{\raggedleft\arraybackslash}p{#1}}

\biboptions{square,numbers,sort&compress}
\graphicspath{{C:/Users/severing/Desktop/HAXPES/manuscript/22_09_2015/plots/}}

\journal{Journal of Electron Spectroscopy and Related Phenomena}

\begin{document}
\begin{frontmatter}

\title{Quantitative study of the $f$ occupation in Ce$M$In$_5$ and other\\ cerium compounds with hard x-rays}

\address[Cologne]{Institute of Physics II, University of Cologne, Z{\"u}lpicher Stra{\ss}e 77, 50937 Cologne, Germany}
\address[Dresden]{Max Planck Institute for Chemical Physics of Solids, N{\"o}thnizer Stra{\ss}e 40, 01187 Dresden, Germany}
\address[NSRRC]{National Synchrotron Radiation Research Center, 101 Hsin-Ann Road, Hsinchu 30077, Taiwan}
\address[RIKEN]{RIKEN SPring-8 Center, 1-1-1 Kouto, Sayo, Hyogo 679-5148, Japan}
\address[SPring-8]{Japan Atomic Energy Agency, SPring-8, 1-1-1 Kouto, Mikazuki, Sayo, Hyogo 679-5148, Japan}
\address[Kwansei]{Graduate School of Science and Technology, Kwansei Gakuin University, Sanda, Hyogo 669-1337, Japan}
\address[LA]{Los Alamos National Laboratory, Los Alamos, NM 87545, US}
\address[CNRS]{Institut NEEL, CNRS, 25 rue des Martyrs, 38042 Grenoble cedex 9, France}
\address[Izumi]{Faculty of Engineering, Toyama Prefectural University, Izumi 939-0398, Japan}
\address[Hiroshima1]{Department of Quantum Matter, AdSM, Hiroshima University, Higashi-Hiroshima 739-8530, Japan}
\address[Hiroshima2]{Institute for Advanced Materials Research, Hiroshima University, Higashi-Hiroshima 739-8530, Japan}

\author[Cologne]{M.~Sundermann\corref{label1}}
\ead{sundermann@ph2.uni-koeln.de}
\cortext[label1]{Corresponding author. Tel.: +49 (351) 4646 - 4323.}
\author[Cologne]{F.~Strigari}
\author[Cologne]{T.~Willers}
\author[Dresden]{J.~Weinen}
\author[NSRRC]{Y.~F.~Liao}
\author[NSRRC]{K.-D.~Tsuei}
\author[NSRRC]{N.~Hiraoka}
\author[NSRRC]{H.Ishii}
\author[RIKEN]{H.~Yamaoka}
\author[Kwansei,SPring-8]{J.~Mizuki}
\author[Kwansei]{Y.~Zekko}
\author[LA]{E.~D.~Bauer}
\author[LA]{J.~L.~Sarrao}
\author[LA]{J.~D.~Thompson}
\author[CNRS]{P.~Lejay}
\author[Izumi]{Y.~Muro}
\author[Hiroshima1]{K.~Yutani}
\author[Hiroshima1,Hiroshima2]{T.~Takabatake}
\author[Hiroshima1]{A.~Tanaka}
\author[Dresden]{N.~Hollmann}
\author[Dresden]{L.~H.~Tjeng}
\author[Cologne]{A.~Severing\corref{label2}}
\ead{severing@ph2.uni-koeln.de}
\cortext[label2]{Corresponding author. Tel.: +49 (221) 470 - 2608.}

\begin{abstract}
We present bulk-sensitive hard x-ray photoelectron spectroscopy (HAXPES) data of the Ce$3d$ core levels and lifetime-reduced $L$-edge x-ray absorption spectroscopy (XAS) in the partial fluorescence yield (PFY) mode of the Ce$M$In$_5$ family with $M$ = Co, Rh, and Ir. The HAXPES data are analyzed quantitatively with a combination of full multiplet and configuration interaction model which allows correcting for the strong plasmons in the Ce$M$In$_5$ HAXPES data, and reliable weights $w_n$ of the different $f^n$ contributions in the ground state are determined. The Ce$M$In$_5$ results are compared to HAXPES data of other heavy fermion compounds and a systematic decrease of the hybridization strength $V_\textrm{eff}$ from CePd$_3$ to CeRh$_3$B$_2$ to CeRu$_2$Si$_2$ is observed, while it is smallest for the three Ce$M$In$_5$ compounds. The $f$-occupation, however, increases in the same sequence and is close to one for the Ce$M$In$_5$ family. The PFY-XAS data confirm an identical $f$-occupation in the three Ce$M$In$_5$ compounds and a phenomenological fit to these PFY-XAS data combined with a configuration interaction model yields consistent results. 
\end{abstract}


\end{frontmatter}

\section{Introduction}
In intermetallic cerium compounds the hybridization of Ce $f$ and conduction electrons ($cf$-hybridization) leads to two competing interactions: the Ruderman-Kittle-Kasuya-Yosida (RKKY) interaction prevailing for weak exchange interactions and favoring a magnetically ordered ground state and the Kondo screening which leads to a nonmagnetic ground state. Non-Fermi liquid behavior and unconventional superconductivity are often observed when going from the verge of one regime to the other. In the presence of strong $cf$-hybridization the $f$-electrons are even partially delocalized. The competition of both interactions can be influenced by pressure, magnetic field or substitution\,\cite{Gegenwart_2008}.

The Ce$M$In$_5$ compounds with $M$\,=\,Co, Rh and Ir are an interesting \textsl{model} system for investigating systematically why a compound orders magnetically or shows unconventional superconductivity because their phase diagram covers a variety of ground states: unconventional superconductivity ($M$\,=\,Co and Ir), antiferromagnetic order ($M$\,=\,Rh), and the coexistence of both e.g. when substituting on the transition metal site. There are also quantum critical points with Fermi surface changes when going from the more itinerant, superconducting Co-rich to the more localized, magnetically ordered Rh-rich side of the substitution phase diagram or when applying pressure to CeRhIn$_5$. It is believed that stronger $cf$-hybridization favors the superconducting ground state over the magnetically ordered one\,\cite{Hegger_2000,Zapf_2001,Petrovic_2001,Petrovic_2001a,Pagliuso_2001,Pagliuso_2002a,Llobet_2005,Park_2006,Ohira-Kawamura_2007, Thompson_2012, Aynajian_2012, Haga_2001,Fujimori_2003,Harrison_2004,Shishido_2005,Settai2007, Shishido, PhysRevLett.101.056402, Allen2013}. It is therefore of interest to quantify the $cf$-hybridization. The $4f$-shell occupation is often used as an indication for the degree of hybridization since strong hybridization favours the delocalization of $f$ electrons. Here, we use a consistent analysis procedure that includes a quantitative plasmon correction to obtain the 4f occupation and that allows a meaningful comparison for materials with very different degrees of hybridization.

In the presence of strong $cf$-hybridization the cerium ground state is no longer a pure $f^1$ state with an $f$-electron count $n_f$\,=\,1. Instead it can be written as a mixed state 
$|\Psi_{\textrm{GS}} \rangle$ = $c_0$~$|f^0\rangle$~+~$c_1$~$|f^1$\underline{L}$\rangle$~+~$c_2$~$|f^2$\underline{\underline{L}}$\rangle$ 
with additional contributions of the divalent and tetravalent states ($f^2$ and $f^0$) and a total $f$ occupation $n_f$ that can be related to the weights $w_n$\,$\equiv$\,$|c_n|^2$ ($n$\,=\,0,1,2) of these ground states as $n_f$\,=\,$w_1$\,+\,2$w_2$. Here \underline{L} and \underline{\underline{L}} denote the conduction band with one and two holes, respectively. The amount of $f^0$ quantifies the degree of delocalization. Photoelectron spectroscopy (PES) and x-ray absorption (XAS) are invaluable in determining such mixed ground states since these -- typically $3d$ core level PES and $L$-edge XAS -- exhibit spectral weights at the energies of corel-electron removal states corresponding to three of these $f$ states: $\underline{c}f^0$, $\underline{c}f^1\underline{L}$ and $\underline{c}f^2\underline{\underline{L}}$, where $\underline{c}$ represents a core hole. Although the intensities $I(\underline{c}f^0)$, $I(\underline{c}f^1\underline{L})$ and $I(\underline{c}f^2\underline{\underline{L}})$ (referred to as $I(f^0)$, $I(f^1)$ and $I(f^2)$ for simplicity) of these structures can be related to the weights of the $f$ states $w_0$, $w_1$, and $w_2$ in the initial state, the relation is not simple because of the effects of the core-hole potential and hybridization in the final state\,\cite{Fuggle1983b,Gunnarsson1983,KotaniJo1988,Gunnarsson2001}. The Anderson impurity model (AIM) in the formalism of Gunnarson and Sch\"onhammer\,\cite{Gunnarsson1983} has been a very successful analysis tool to relate final state spectral weights $I(f^n)$ to the respective $f$ contributions $w_n$ in the initial (ground) state. In Ref.\,\cite{Fuggle1983b} Fuggle \textsl{et al.} show and analyze an impressive amount of PES data of Ce compounds giving number for $w_0$ and hybridization parameters. However, PES data are often subject to surface effects and the surface valence is usually closer to integer than the one of the bulk\,\cite{Laubschat1990}. Furthermore, the AIM in combination with a full multiplet routine would require elaborate computation so that usually the spectral weights are assigned phenomenologically. However, the spectral shapes (energy distribution) of the $f^1$ and $f^2$ contributions depend on the hybridization so that the assignment of spectral weights to $I(f^n)$ is not trivial, something desperately needed when aiming at a quantitative plasmon correction. The non-trivial assignment of spectral weights is also valid for $L$-edge XAS where the spectral shapes are determined by the empty $5d$ density-of-states ($5d$-DOS). This is valid even in the state-of-the-art partial fluorescent yield mode (PFY) which gives a much better contrast than conventional total fluorescence yield (TFY)-XAS\,\cite{RIXS_1991,Kotani2001,Dallera2002,Rueff2010,Zekko2014}. It is therefore not trivial to find comparable numbers for the $f$-electron occupation.

There have been several attempts to determine the $4f$-occupation of the Ce$M$In$_5$ family, however, quantitatively they are not conclusive\,\cite{Fujimori_2003,Daniel_2005,Fujimori_2006,Gam_2008,Willers_2010,Booth_2011,Treske2014}. XAS data at the Ce $M$-edge or resonant data at the $N$-edge of the Ce$M$In$_5$ compounds show only minor amounts of $I(f^0)$ without further quantification\,\cite{Fujimori_2003,Fujimori_2006,Willers_2010}. A more adventurous attempt to quantify the $f$-occupation of CeCoIn$_5$ and CeCo(In$_{0.85}$Cd$_{0.15}$)$_5$ in an $M$-edge XAS experiment was made by Howald \textsl{et al.} \cite{Howald_2015} from an extrapolation of CeF$_3$ and CeO$_2$ $M$-edge data, assuming the former is tri- and the latter tetravalent. However, there is a lot of work showing that CeO$_2$ appears strongly covalent ($n_f$\,$\approx$\,0.5) in core-level spectroscopies like $L$- or $M$-edge XAS or PES (see e.g. Ref.'s\,\cite{Kotani1985a,Kotani1985b,Hague_2004,Kotani_2011} and references therein), thus putting into question the $f$-occupation of 0.85 resulting from this analysis. TFY-XAS data at the Ce $L$-edge yield $n_f$\,=\,0.9for CeCoIn$_5$ \,\cite{Booth_2011}, and 0.98 and 0.96 for the Rh and Ir compounds, respectively\,\cite{Daniel_2005}. In these TFY $L_3$-edge XAS data the $I(f^n)$ weights are not well resolved. The assignment is further complicated by the fine structure of the empty $5d$-DOS and the extended x-ray absorption fine structures (EXAFS) which in these compounds are close in energy to the main absorption line\,\cite{Daniel_2005,Booth_2011}. PES data at the Ce$3d$ core level were taken with Al K$_{\alpha}$ radiation\,\cite{Gam_2008,Treske2014}. They bear the problem of surface sensitivity\,\cite{Laubschat1990} and in addition exhibit strong plasmons at the energy of the $I(f^0)$ spectral weight. $n_f$\,=\,0.9 was given as a lower limit for all three compounds without correcting quantitatively for the plasmons\,\cite{Treske2014}.

Here we present a quantitative analysis of bulk-sensitive HAXPES data of the Ce$M$In$_5$ family and compare data and analysis with the ones of the large T$_K$ ($\approx$ 600\,K) and intermediate valent compound CePd$_3$\,\cite{Schneider_1981,Bianconi_1981,Fuggle1983b,Croft_1984,KotaniJo1988,Murani_1996,Fanelli_2014}, intermediate valent and ferromagnetically ordered CeRh$_3$B$_2$ ($T_c$\,=\,115\,K)\,\cite{Dhar_1981,Sampathkumaran_1985,Fujimori_1990} and the heavy fermion compound CeRu$_2$Si$_2$, which does not order magnetically at ambient pressure and zero magnetic field\,\cite{Flouquet_2005,Aoki2014}. 

Our HAXPES data analysis comprises a quantitative plasmon correction which we achieve by combining a full multiplet (fm) calculation with a configuration interaction (CI) model (fm-CI) with a single-state-approximation for the conduction band. It is a convenient simplification to determine the $I(f^n)$ spectral weights and relate them to $w_n$ weights in the ground state. The model was already suggested by Imer and Wuilloud\,\cite{Imer1987} and, although it has certain drawbacks\,\cite{Imer1987,Strigari2015}, has the great advantage that it can be combined with a full multiplet calculation. This is particularly important for the Ce$M$In$_5$ compounds where strong plasmons would otherwise hamper the quantitative determination of spectral intensities\,\cite{Gam_2008,Treske2014}. The full multiplet calculation offers the possibility to describe the plasmon intensities as part of the line shape of each emission line, after having determined the plasmon parameters from core levels that are not affected by the configuration interaction (In$3p$ for each Ce$M$In$_5$, Pd$3p$ for CePd$_3$, Rh$3d$ for CeRh$_3$B$_2$, and Ru$3d$ for CeRu$_2$Si$_2$). We have used this type of analysis already successfully for the Ce$T_2$Al$_{10}$ compounds\,\cite{Strigari2015} and we apply it here to obtain quantitative $f$-occupations for the Ce$M$In$_5$ compounds. The resulting values will be compared with the ones of CePd$_3$, CeRh$_3$B$_2$, and CeRu$_2$Si$_2$. The model is described by the CI parameters for the Coulomb exchange interaction between the $f$ electrons ($U_{ff}$) and between the $f$ electrons and $3d$ core hole ($U_{fc}$), the effective $f$-electron binding energy $\Delta_f$ and the isotropic hybridization $V_\textrm{eff}$. The energy distances and intensities of the three $I(f^n)$ spectral weights in the $3d$ core level HAXPES data yield sufficient information for determining the four CI parameters in a unique manner. 

We confirm our findings by showing $L$-edge absorption data in the PFY mode where a decay process with longer life time is selected so that the life time broadening of the XAS spectra can be reduced, thus facilitating the separation of the different $I(f^n)$ spectral weights\,\cite{RIXS_1991,Kotani2001,Dallera2002,Rueff2010}. Here we compare the Ce$M$In$_5$ spectra with the ones of the more strongly hybridized compounds Ce$T$Al$_{10}$\,\cite{Strigari2015} and present a fit to the data which is consistent with the HAXPES results.

The experimental set-ups of HAXPES and PFY-XAS including details for sample preparation are given in Appendix A and information of the data simulation and HAXPES line shapes are given in Appendix B. 

\begin{figure}
   \centering
   \includegraphics[width=1.00\columnwidth]{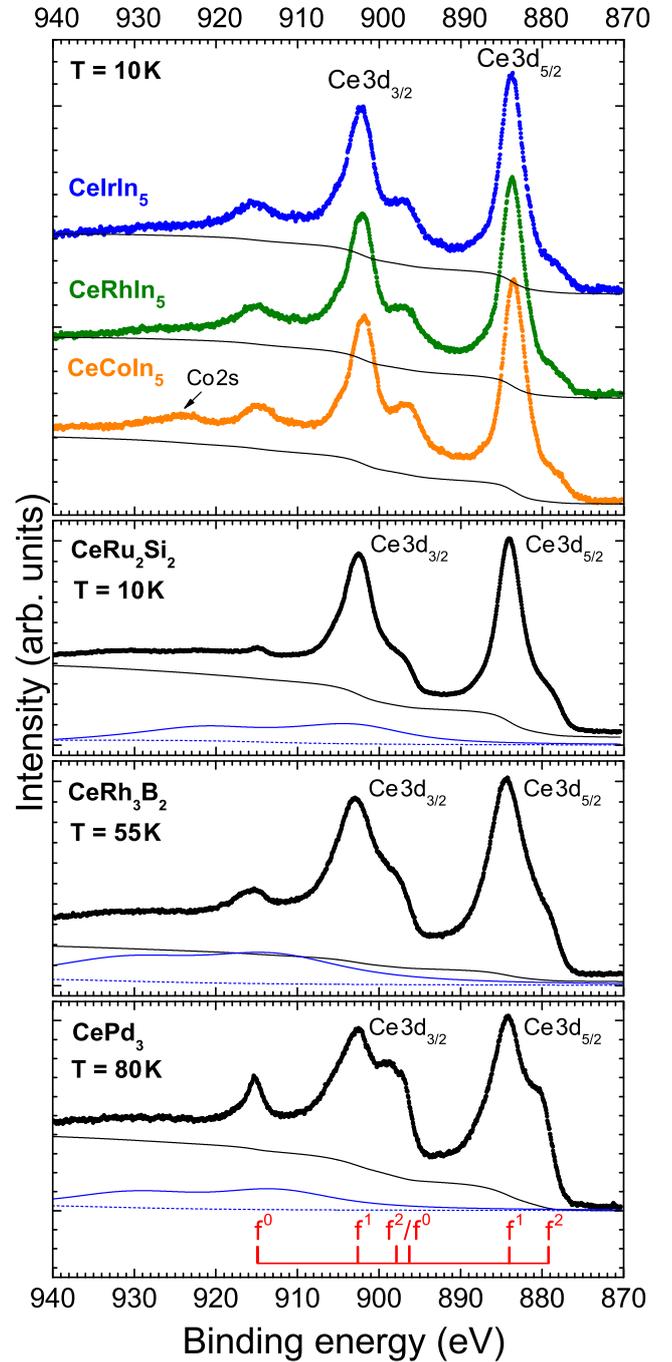}
    \caption{Low temperature Ce$3d$ HAXPES data as measured of Ce$M$In$_5$ ($M$\,=\,Ir, Rh, and Co), CeRu$_2$Si$_2$, CeRh$_3$B$_2$, and CePd$_3$. The solid black lines show the respective integral background. The solid and dashed blue lines show the 1$^\textrm{st}$ and 2$^\textrm{nd}$ plasmon contribution (see text). For the Ce$M$In$_5$ this is shown in Figure 3. The red ruler in the bottom panel indicates the energy positions of the $I(f^n)$ spectral weights.}
    \label{data}
\end{figure}

\section{Data and simulation} 
Figure\,\ref{data} shows the uncorrected $3d$ core level HAXPES data of Ce$M$In$_5$ ($M$\,=\,Ir, Rh, and Co), CeRu$_2$Si$_2$, CeRh$_3$B$_2$, and CePd$_3$. All spectra exhibit two sets of emission lines, due to the Ce$3d$ spin orbit splitting. Each set, i.e$.$ the Ce$3d_{3/2}$ and the Ce$3d_{5/2}$, contains three spectral weights because of the core hole effect on the mixed ground state. At the bottom of the CePd$_3$ data the energy positions of the two sets of $I(f^0)$, $I(f^1)$, and $I(f^2)$ are marked. It shows that the expected $I(f^0)$ spectral weight at 915\,eV binding energy does not overlap with any other spectral feature and that the $I(f^2)$ intensity at $\approx$880\,eV overlaps only partially with $I(f^1)$. However, there is a multiplet structure underneath $I(f^1)$ and $I(f^2)$ which has to be accounted for when disentangling the two. 

The Ce$M$In$_5$ data in Fig.\,\ref{data} look very much alike. Only the Co data exhibit an extra hump at about 923 eV binding energy which is due to some intensity from the Co$2s$ emission. We now compare qualitatively the ratio of the $I(f^2)$/$I(f^1)$ intensities for the Ce3d$_{5/2}$ emission: the ratio increases systematically from top to bottom, i.e$.$ from the Ce$M$In$_5$ family to CePd$_3$. 
However the intensity ratio $I(f^0)$/$I(f^1)$ seemingly does not have the same systematic when comparing the intensities at the $I(f^0)$ position at 915\,eV. For the Ce$M$In$_5$ $I(f^0)$/$I(f^1)$ seems larger than for CeRu$_2$Si$_2$. This is puzzling since increasing hybridization should lead to an increase of both $I(f^2)$/$I(f^1)$ and $I(f^0)$/$I(f^1)$. 

\begin{figure}
   \centering
   \includegraphics[width=0.96\columnwidth]{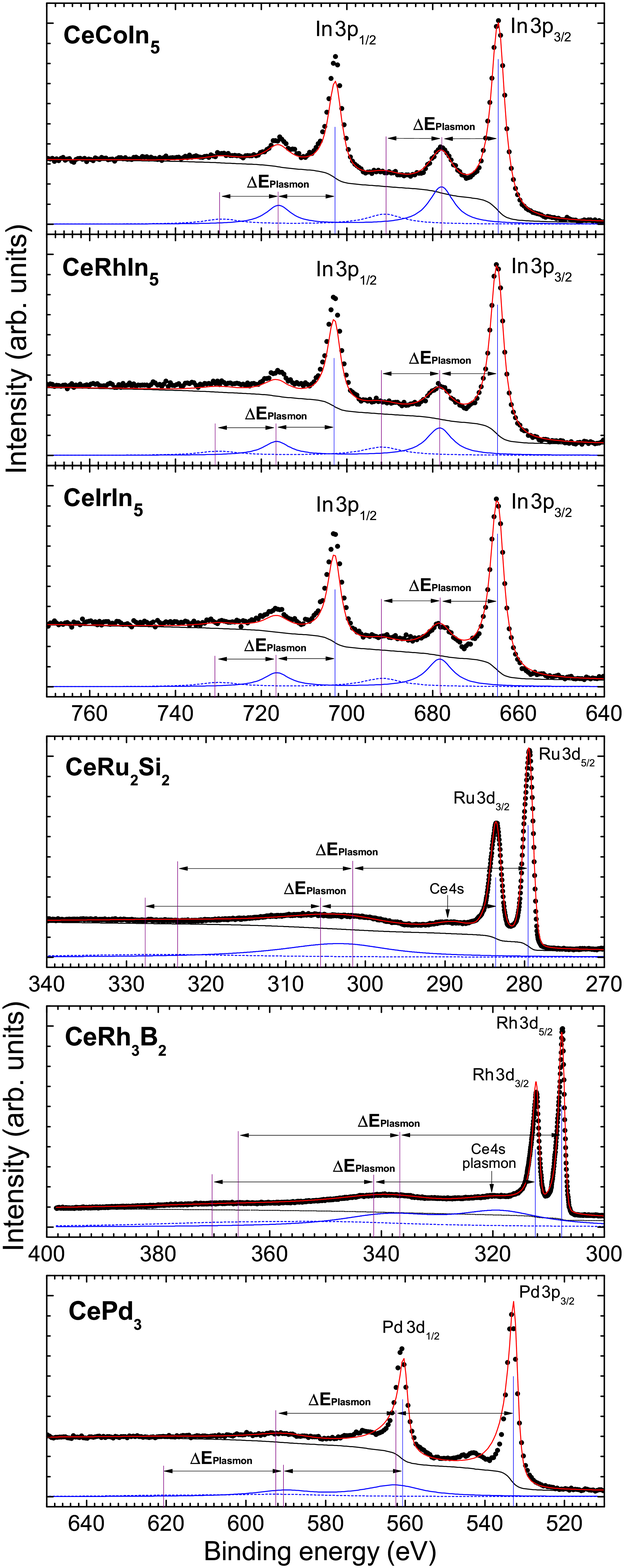}
     \caption{Emission lines as measured of core levels that are not affected by the configuration interaction. The solid black lines are the integral background, the red lines are the result of an ionic calculation with adequate line shape parameters. The blue lines (solid and dashed) represent the total (1$^\textrm{st}$ and 2$^\textrm{nd}$ order) plasmon intensities which were obtained by fitting each emission line with a line shape consisting of the main emission line plus a 1$^\textrm{st}$ and 2$^\textrm{nd}$ plasmon contribution at equidistant energies.}
    \label{plasmons}
\end{figure}

This puzzle is quickly solved when looking at the In$3p$ emission spectra of the Ce$M$In$_5$ compounds at the top of Fig.\,\ref{plasmons}. Strong and relatively sharp intensities in addition to the main In$3p_{1/2}$ and In$3p_{3/2}$ emission lines are visible at about 13-14\,eV higher binding energies, respectively, and they are identified as plasmonic excitations. This is about the energy distance of the $I(f^0)$ and $I(f^1)$ spectral weights in the Ce$3d$ core level spectra, meaning the $I(f^0)$ intensities in the Ce$M$In$_5$ data are superimposed by plasmons as already pointed out by the authors of Refs.\,\cite{Gam_2008,Treske2014}, leading to the misleading impression that the $I(f^0)$ spectral weight in the 115 compounds is larger than in CeRu$_2$Si$_2$. 

\begin{figure}
\centering
    \includegraphics[width=1.00\columnwidth]{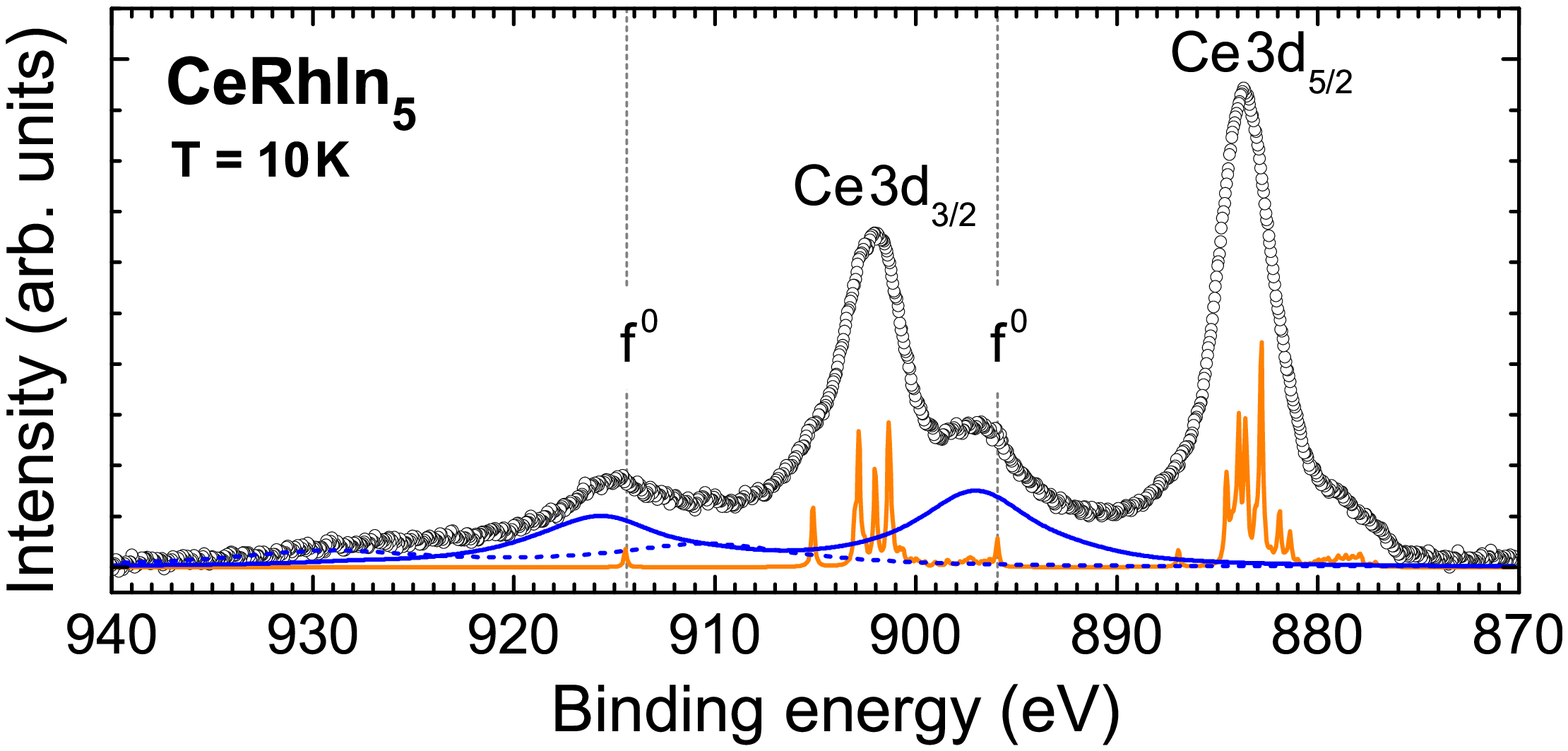} 
    \caption{Ce$3d$ core level emission data of CeRhIn$_5$ after subtraction of the integral background. The blue lines (solid and dashed) represent the total 1$^\textrm{st}$ and 2$^\textrm{nd}$ order plasmon intensities which were obtained by fitting each emission line of the multiplet structure (orange, times 0.1) with a line shape consisting of a main emission line plus a 1$^\textrm{st}$ and 2$^\textrm{nd}$ plasmon contribution at equidistant energies. The vertical gray lines indicate the single emission line of $I(f^0)$.}
    \label{CeRhIn5_In3p_plasmon}
\end{figure}

 \begin{table*}
   \centering
    \caption{Top: Optimized configuration interaction parameters  ($U_{ff}$, $U_{fc}$, $V_\textrm{eff}$, $\Delta_f$) in eV and resulting $f^n$ contributions 
		$w_0$, $w_1$, and $w_2$ in the ground state in \%. The total $f$ electron count is given by $n_f$\,=\,$w_1$\,+\,2$w_2$.} 		
		\label{Tab_simulation}
    \begin{tabular*}{0.98\textwidth}{@{\extracolsep{\fill}}lrrrrrrr}
    \hline \hline                          
                    &        & CePd$_3$     & CeRh$_3$B$_2$  & CeRu$_2$Si$_2$   & CeCoIn$_5$       & CeRhIn$_5$       & CeIrIn$_5$        \\
    \hline
      $U_{ff}$      & [eV]     & 11.4         & 10.2           & 9.2              & 8.5              & 8.5              & 8.5           \\
      $U_{fc}$      & [eV]     & 11.6         & 10.9           & 9.9              & 9.9              & 9.9              & 9.9           \\
      $V_\textrm{eff}$ & [eV]  & 0.36         & 0.31           & 0.25             & 0.20             & 0.20             & 0.20          \\
      $\Delta_{f}$  & [eV]     & 1.3          & 2.0            & 2.6              & 2.4              & 2.4              & 2.4            \\
		\\
      $w_0$          & [\%]    & 24.0 ($\pm$3.0) & 13.5 ($\pm$3.0)&  6.8 ($\pm$1.2)  &  5.0 ($\pm$1.5)  &  4.7 ($\pm$1.5)  &   4.7 ($\pm$1.5) \\
      $w_1$          & [\%]    & 74.0 ($\pm$3.0) & 83.7 ($\pm$3.0)& 90.3 ($\pm$1.2)  & 93.0 ($\pm$1.5)  & 93.3 ($\pm$1.5)  &  93.3 ($\pm$1.5) \\
      $w_2$          & [\%]    &  2.0 ($\pm$0.5) &  2.6 ($\pm$0.5)&  2.9 ($\pm$0.5)  &  2.0 ($\pm$0.5)  &  2.0 ($\pm$0.5)  &   2.0 ($\pm$0.5) \\ 
	    \hline \hline
  \end{tabular*}
\end{table*}

\begin{figure}
   \centering
   \includegraphics[width=1.00\columnwidth]{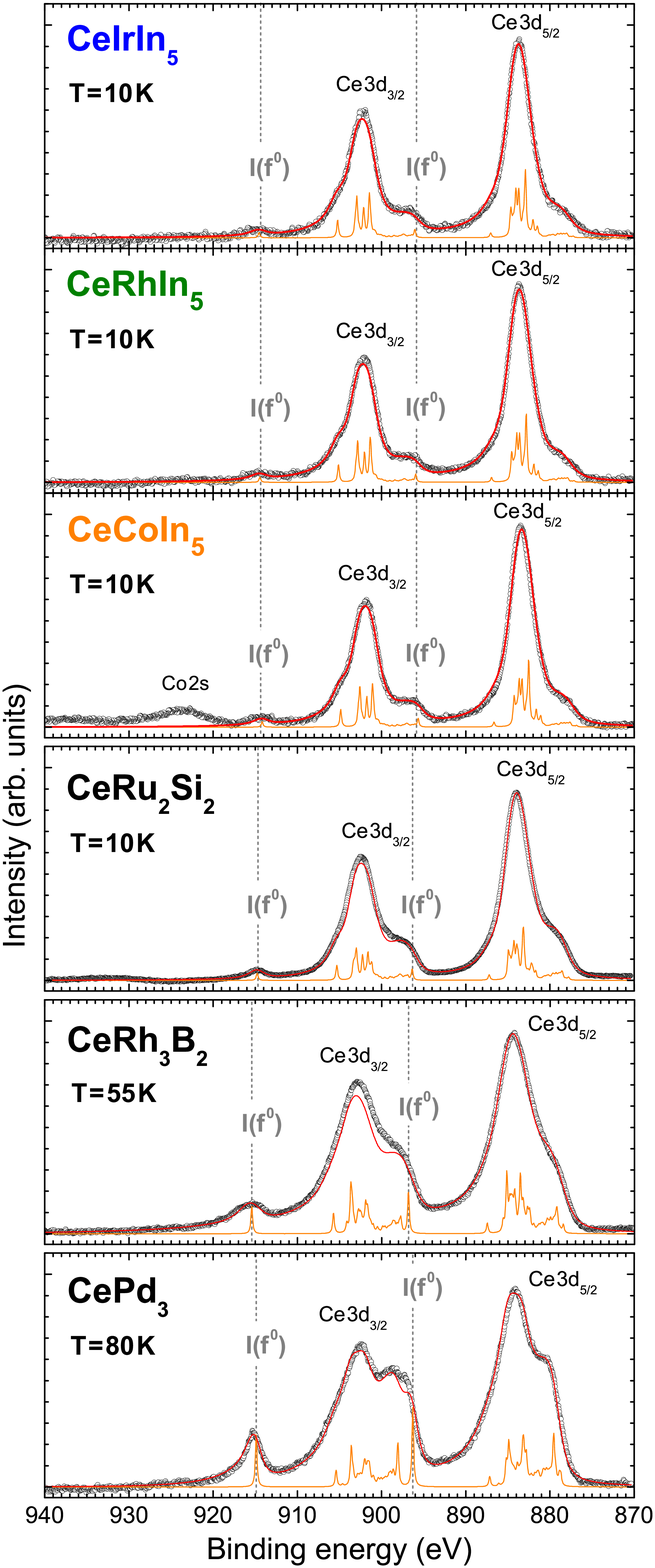}
     \caption{Ce3$d$ emission data after background and plasmon subtraction (black circles). The red lines are the result of the fm-CI calculation (see text), the orange lines resemble the corresponding multiplet structures times 0.1. The vertical gray lines indicate the single emission line of $I(f^0)$.}
    \label{fit}
\end{figure}

The data analysis comprises 1) the background correction, 2) the plasmon correction, 3) the assignment of spectral weights, and 4) the conversion of final state spectral weights $I(f^n)$ into initial state contributions $w_n$ with the CI calculation. Step 1) is straightforward and an integrated background as marked by the black lines in Fig.\,\ref{data} is subtracted from the data before further analysis. In the presence of strong plasmons the steps 2), 3) and 4) have to be performed in one go by using line shapes for the multiplet excitations that consist of the main emission plus a first and second order plasmon. 

The plasmon contributions to the line shapes are determined from fits to the core hole emissions spectra (after background correction) that are not affected by the configuration interaction. These spectra and line-shape fits are shown in Fig.\,\ref{plasmons} for all compounds. The lines are broadened by a Gaussian and Lorentzian function to account for instrumental resolution and lifetime broadening. In addition a Mahan function is used to account for the asymmetry of the line shapes. Then a single and double plasmon excitation is attached to each multiplet line. The plasmon excitations in CeRu$_2$Si$_2$, CeRh$_3$B$_2$, and CePd$_3$ are much broader than in the Ce$M$In$_5$ compounds and appear at larger energy distances from the main emission line. Nevertheless the same procedure was used for all compounds. The resulting plasmon parameters are listed in the Appendix B in Table\,\ref{Tab_lineshape}. 

Having determined the plasmon energies, line widths and intensity ratios, the combined fm-CI is applied to the background-corrected Ce$3d$ data. The plasmon contributions in the Ce$3d$ emission data of the Ce$M$In$_5$ resulting from such a fit are shown exemplary for CeRhIn$_5$ in Fig.\,\ref{CeRhIn5_In3p_plasmon} and for the other compounds in Fig.\,\ref{data} (see solid (1$^\textrm{st}$) and dashed (2$^\textrm{nd}$) blue lines). Finally, Fig.\,\ref{fit} shows the background- and plasmon-corrected data (open circles). Now the ratios $I(f^0)$/$I(f^1)$ and $I(f^2)$/$I(f^1)$ are smallest for the 115 compounds. They increase to CeRu$_2$Si$_2$ and are largest for CePd$_3$. The red lines are the result of the fm-CI simulation. The resulting fit parameters and spectral weights are listed in Table\,\ref{Tab_simulation}. The complete set of lineshape parameters is given in Table\,\ref{Tab_lineshape}. The orange lines indicate the underlying multiplet structures. Note, $I(f^0)$ consists of only one emission line so that it remains visible even when the intensity is rather weak (see vertical gray lines in Fig.\,\ref{fit} marking the sharp peaks in the multiplet structure at $\approx$915 and $\approx$896\,eV). 

\begin{figure}[t]
   \centering
   \includegraphics[width=1.00\columnwidth]{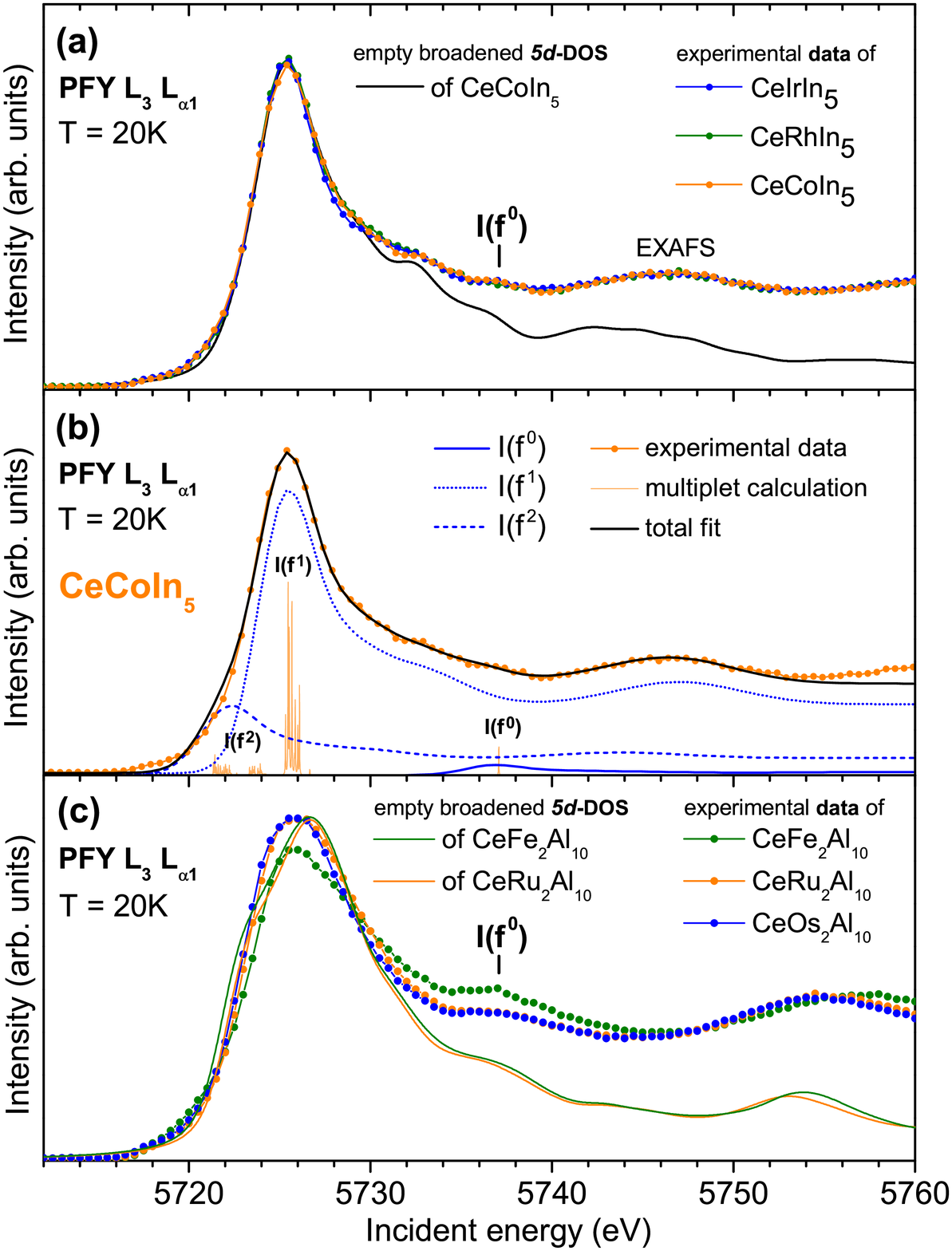}
     \caption{a) $L$-edge PFY-XAS spectra of Ce$M$In$_5$ normalized to the extended x-ray absorption fine structure (EXAFS) beyond 5742\,eV. The black lines resemble Wien2k calculations of the unoccupied $5d$-DOS for CeCoIn$_5$. b) Simulation of CeCoIn$_5$ data with identical line shapes for each $I(f^n)$ based on a CI model with parameters as for HAXPES, only $U_{fc}$ was adjusted (see text). The line shapes are approximated by Gaussians and a step funtion (see text). The orange lines are the underlying multiplet structure of each $I(f^n)$ (see text). c) $L$-edge PFY-XAS data of Ce$T_2$Al$_{10}$ with $T$\,=\,Fe, Ru, and Os plus $5d$-DOS from Wien2k calculations for CeFe$_2$Al$_{10}$ and CeRu$_2$Al$_{10}$.}
		\label{PFY}
\end{figure}

We also show PFY-XAS data where the $L_{\alpha 1}$ decay process $3d_{5/2}$ $\rightarrow$ $2p_{3/2}$ after the $L_3$ absorption $2p$ $\rightarrow$ $5d$ is measured (see Fig.\,\ref{PFY}(a)) in order to confirm our HAXPES result of a minor and very similar $cf$-hybridization in the Ce$M$In$_5$ compounds. Also here we find that at 20\,K the spectra of all three compounds are identical within the accuracy of the measurement. It is actually next to impossible to spot the $I(f^0)$ spectral weight also when comparing the data at 20\,K with data taken at 200\,K (not shown). Band structure calculations of the empty $5d$-DOS neglecting core hole effects (see Appendix C) show that the structures at 5733 and 5746\,eV are due to band structure (see dashed and solid black lines in Fig.\ref{PFY}(a)). However, comparing the PFY-XAS data of the Ce115s with the PFY-XAS data of the more strongly hybridized Ce$T_2$Al$_{10}$ compounds quickly shows where the $I(f^0)$ spectral weight should be. CeFe$_2$Al$_{10}$ is known to be more strongly intermediate-valent than the $T$\,=\,Ru and Os samples\,\cite{Strigari2015} and there is indeed a large difference in the spectra. The strongest absorption line at 5726\,eV is strongly reduced while the intensity at 5737\,eV is enhanced, i.e$.$ the latter must be the $I(f^0)$ spectral weight. Also here the band structure calculations reproduce the overall features of the the EXAFS and the similarity of the $5d$-DOS of CeFe$_2$Al$_{10}$ (green line) and CeRu$_2$Al$_{10}$ (orange line) in Fig.\,\ref{PFY}(c) confirms that the difference in the data must indeed be due to the difference in $I(f^n)$ spectral weights. The respective central energy positions of the $I(f^n)$ are marked in Fig.\ref{PFY}(a). 

In Fig.\,\ref{PFY}(b) we show a consistency fit to the PFY data of Ce$M$In$_5$. Applying a CI calculation using the same initial state Hamiltonian as for HAXPES yields also good results for the PFY-XAS data when using identical lineshapes for each $I(f^n)$, consisting of three Gaussians, a tanh-type step function with inflection point nearby the maximum of the respective $I(f^n)$ absorption, and an additional Gaussian for mimicking the EXAFS. This choice of lineshape has actual similarities with the empty $5d$-DOS in Fig.\,\ref{PFY}(a). Here the weighted, averaged energy distributions of the multiplet states define the energy separations of the respective $I(f^n)$. The red line resembles the total fit. Here only the CI parameter $U_{fc}$ has been adjusted from 9.9 as for HAXPES to 9\,eV for PFY-XAS in order to match the energy position of $I(f^0)$. Please note that the same adjustment is necessary for Ce$T_2$Al$_{10}$ where HAXPES yields an $U_{fc}$ of 10\,eV \cite{Strigari2015} while the present PFY-XAS data require 9\,eV. The reason is most likely the screening of the $2p$ hole in the XAS final state by the additional $5d$ electron. 

\section{Results and Discussion}
The present results shall be briefly compared with some of the data available in literature before discussing the systematic of the resulting CI parameters in Table\,\ref{Tab_simulation}. There are several electron spectroscopy measurements of the valence of CePd$_3$. To name some, TFY $L_3$-edge absorption yields values of about $\approx$3.15\,\cite{Bianconi_1981,Croft_1984} and PES data by Fuggle \textsl{et al.}\,\cite{Fuggle1983b} and Kotani \textsl{et al.}\,\cite{KotaniJo1988} yield an $f^0$ contribution of $w_0$\,=\,10\%  and an $f$-occupation of $n_f$\,=\,0.86, respectively. The last two works take final state effects into account. We find a stronger deviation from an integer $f$-occupation which might be due to the fact that the before mentioned PES data suffer from surface effects which result in an $f$ occupation closer to integer\,\cite{Laubschat1990}.

Also in CeRh$_3$B$_2$ the presence of an important amount of $f^0$ in the $L$-edge XAS data was reported\,\cite{Sampathkumaran_1985} and an $f$ occupation of $n_f$\,=\,0.85 was given from the quantitative analysis of $3d$ core level PES data by Fujimori \textsl{et al.}\,\cite{Fujimori_1990} The PES value is close to what we obtain from the bulk-sensitive data and we can only speculate why we find a slightly \textsl{smaller} deviation from integer valence in the bulk-sensitive data. Here surface effects are no explanation since they rather increase the $f$-occupation\,\cite{Laubschat1990}. However, our total fm-CI fit to the CeRh$_3$B$_2$ data is not that perfect and this might be due to the giant crystal-electric field (CEF) in this compound\,\cite{Givord2007}. In contrast to Fujimori \textsl{et al.}\,\cite{Fujimori_1990}, we used a single crystal for the HAXPES experiment and possible polarization dependencies due to CEF effects as recently reported\,\cite{Mori2014} are not part of our calculation. Hence there is room for some error. 

For CeRu$_2$Si$_2$ some bulk-sensitive HAXPES data, analyzed with an AIM are available\,\cite{Suga2008}. Here the valence band continuum was divided into $N$\,=\,21 discrete levels and the spectral weights were assigned empirically. The $f^0$ occupation is almost the same as from our analysis, but the CI-parameters differ and the empirical assignment of $I(f^2)$ seems to lead to an overestimation of the latter ($n_f$\,=\,0.987, $w_0$=\,5\%, and $w_2$\,=\,4.7\% according to Yano \textsl{et al.}\,\cite{Suga2008}). 

The values in Table\,\ref{Tab_simulation} show the largest values for $U_{ff}$ and $U_{fc}$ for the most strongly hybridized compounds CePd$_3$. For the more weakly hybridized compounds the fits were less sensitive to $U_{fc}$ so that it was kept at a constant value of 9.9\,eV. The effective hybridization $V_\textrm{eff}$ is clearly largest for intermediate valent CePd$_3$ and smallest for the Ce$M$In$_5$ compounds while their effective $f$ electron binding energy $\Delta_f$ is comparable to that CeRu$_2$Si$_2$. The resulting $f^n$ contributions in the ground state wave function are also given in Table\,\ref{Tab_simulation} and we can clearly state that the Ce$M$In$_5$ compounds have the smallest $f^0$ contribution in the ground state, even less than CeRu$_2$Si$_2$.  Interestingly, the $f^2$ contribution in the ground state ($w_2$) is almost the same (about 2-3\%) for \textsl{all} compounds, from strongly intermediate valent CePd$_3$ to the much less hybridized Ce$M$In$_5$, emphasizing the importance of final state effects. In the final state, $f^1$ and $f^2$ are close in energy and therefore strongly entangled. As a result, the spectral weights $I(f^2)$ may be fairly strong in HAXPES as well as PFY-XAS. 

Within the accuracy of the present HAXPES and PFY-XAS data, the different ground states in the Ce$M$In$_5$ cannot be due to differences in the $f$ shell occupation. However, the superconducting compounds CeIrIn$_5$ and CeCoIn$_5$ show enlarged Fermi surface volumes, implying a more delocalized $f$-electron behavior\,\cite{Haga_2001,Fujimori_2003,Harrison_2004,Shishido_2005,Settai2007,Allen2013}, especially in the purely superconducting regions of the substitution phase diagrams\,\cite{Shishido,PhysRevLett.101.056402}, while in magnetically ordered CeRhIn$_5$ the $f$ electrons do not contribute to the Fermi surface, i.e remain localized. These findings are supported by inelastic neutron measurements of the quasielastic line width; it is seems much broader for the Co and Ir samples with respect to the Rh one\,\cite {Willers_2010}. In order to rationalize these experimental results of identical $f$-occupations but different Fermi surfaces we apparently have to look beyond the idea of an isotropic hybridization that leads to a measurable delocalization. In Ref.\,\cite{PNAS_2015} was shown for CeRh$_{1-x}$Ir$_x$In$_5$ that the local $f$ orbital symmetry, i.e. the CEF ground state wave function is an important parameter for the ground state formation. According to this experimental study, taller $f$ ground state orbitals favor a superconducting ground state. These findings agree with first-principle dynamical mean field theory calculations by Shim \textsl{et al.}\,\cite{Shim_2007} which postulated for CeIrIn$_5$ that the out-of-plane hybridization with the so-called In(2) is the most important one. Hence, it is not the absolute strength of the $cf$-hybridization (and occupation) that changes among the Ce$M$In$_5$ compounds, it rather seems to be the efficiency of hybridization that depends on the orbital anisotropy.

\section{Summary}
The $f$-occupation of the Ce$M$In$_5$ family has been investigated with bulk-sensitive HAXPES and PFY-XAS and compared with spectra of cerium compounds of different hybridization strengths. The PFY-XAS data show that the $f$-occupation of the Ce$M$In$_5$ compounds is identical within the accuracy of the present experiments. A detailed analysis of the HAXPES data, which includes a quantitative plasmon correction, shows further that the effective hybridization $V_\textrm{eff}$ and also the amount of $f^0$ in the ground state in the Ce$M$In$_5$ is even smaller than in the non-magnetically ordering compound CeRu$_2$Si$_2$. 

\section*{Acknowledgments}
For M.S., F.S. and A.S. this work was supported by DFG through project 600575. We thank S. Wirth for fruitful discussions and reading the manuscript. The PFY experiment was performed under the approvals with Japan Synchrotron Radiation Research Institute (Proposal No. 2011A4254, 2012A4252, 2012B4251) and National Synchrotron Radiation Research Center, Taiwan (2011-2-053, 2012-3-078). Work at Los Alamos was performed under the auspices of the U.S. Department of Energy, Office of Basic Energy Sciences, Division of Materials Sciences and Engineering.

\section*{Appendix A. Experimental set-ups}
The HAXPES spectra of the Ce\,3$d$ emission were taken at the Taiwan beamline BL12XU at SPring-8 with an incident energy of 6.5\,keV and horizontally polarized light. An MB Scientific A1-HE analyzer at an angle of 90$^\circ$ to the incident beam in the vertical (horizontal for CeRh$_3$B$_2$) plane gives an overall instrumental resolution of $\approx$1\,eV at the Ce 3$d$ emission using a pass energy of 200\,eV and fully opened slits (S3.2). The Fermi energy and instrumental resolution was obtained by measuring the valence band spectra of silver or gold thin films. 
Clean sample surfaces were obtained by cleaving \textsl{in situ} under ultra high vacuum of the order of 10$^{-9}$\,mbar and the reproducibility of successively taken scans showed the absence of aging effects. Only CeRh$_3$B$_2$ showed one additional line at C$1s$ energy, indicating some graphite on the surface, which may also explain the larger broadening in this compound. For the other samples, scans over a wide energy range, from the Fermi energy to 1\,keV binding energy, verified the absence of impurities.

The PFY-XAS experiments were performed at the inelastic scattering beamline with a Johann-type set-up at the synchrotron SPring-8 in Japan (BL12XU beamline). The undulator beamline uses a cryogenically-cooled double crystal monochromator to obtain a monochromatic beam. The $3d_{5/2}$ $\rightarrow$ $2p_{3/2}$ de-excitation of 4800 eV at the Ce $L_3$ absorption edge was measured as function of the incident energy. The emission line was analyzed with a spherically bent Si(400) analyzer crystal (radius 1\,m). The analyzed photons were detected in Si solid state detectors (Amptech) with an overall energy resolutions of $\approx$ 1.5~eV . The intensities of the measured spectra were normalized with a monitor just before the sample. The beam size was 120(h)~$\times$~80(v)~$\mu$m$^2$. The flight path was filled with He to reduce air scattering. The closed-circuit He cryostat reached a base temperature of 20\,K.

The Ce$M$In$_5$ single crystals were grown with the flux growth method, the CeRu$_2$Si$_2$ and the CeRh$_3$B$_2$ single crystals by the Czochralski technique. The CePd$_3$ polycrystalline sample was made by arc-melting the constituents on a water-cooled Cu hearth under a UHP Argon atmosphere. Polycrystalline samples of Ce$T_2$Al$_{10}$ were prepared by arc-melting the constituents amounting of pure elements under Argon atmosphere and annealing at 850$^{\circ}$\,C for one weak. 

\section*{Appendix B. Simulation}
The data have been analyzed with the combined fm-CI model as described in detail in Ref.\,\cite{Strigari2015}. For the fm-CI simulations the XTLS 9.0 code by A. Tanaka was used\,\cite{TanakaJPSC63} and the atomic input parameters for the intra-atomic $4f$-$4f$ and $3d$-$4f$ Coulomb and exchange interactions and the $3d$ and $4f$ spin-orbit coupling were calculated with Cowan's atomic structure code\,\cite{Cowan}. Reduction factors of $\approx$40\% and $\approx$20\% for the atomic Hartree-Fock values for the $4f$--$4f$ and $3d$--$4f$ Coulomb interaction were used\,\cite{Strigari2015}. The hybridization between the $f$ and conduction electrons is described by the Coulomb exchange interaction between the $f$ electrons ($U_{ff}$) and between the $f$ electrons and $3d$ core hole ($U_{fc}$), the effective $f$-electron binding energy $\Delta_f$ (i.e$.$ the energy difference between the $f^0$ and $f^1\underline{L}$ in the initial state) and the isotropic hybridization $V_\textrm{eff}$. 

Plasmons appear at well-defined energies at higher binding energies so that the full multiplet calculations allows the the pinning of a plasmon and its multiples to each emission line with the same energy distance, line width and shape. The line shape parameters for life time broadening (Lorentzian $\&$ Mahan) and the energies and relative intensities of 1$^\textrm{st}$ and 2$^\textrm{nd}$ plasmons were determined from fitting the core level spectra of another element in the sample which is not affected by the CI\,\cite{Strigari2015}). We used In\,$3p$ for the Ce$M$In$_5$ compounds, Pd\,$3p$ for CePd$_3$, Rh\,$3d$ for CeRh$_3$B$_2$, Ru\,$3d$ for CeRu$_2$Si$_2$. The line shape parameters which were used in the data analysis are listed in Table\,\ref{Tab_lineshape}. 

    \begin{table*}
   \centering
    \caption{Line shape parameters of the Ce$3d$ emission lines; Gaussian FWHM$_G$, Lorentzian FWHM$_L$, and the Mahan broadening with asymmetry $\alpha_M$ and cut-off $\gamma_M$, plus the plasmon energy $\Delta$E$^{Pl}$ and width of the plasmon excitation FWHM$_L^{Pl}$.} 		
		\label{Tab_lineshape}
    \begin{tabular*}{0.98\textwidth}{@{\extracolsep{\fill}}lrrrrrrr}
    \hline \hline                          
                     &       & CePd$_3$        & CeRh$_3$B$_2$   & CeRu$_2$Si$_2$   & CeCoIn$_5$       & CeRhIn$_5$       & CeIrIn$_5$        \\
    \hline
      FWHM$_G$       &  [eV]     & 1.1         & 2.0*            & 1.1              & 1.1              & 1.1              & 1.1           \\
      FWHM$_L$       &   [eV]    & 1.14        & 1.36            & 1.44             & 1.36             & 1.36             & 1.3         \\
      $\alpha_M$     &           & 0.40        & 0.2             & 0.1              & 0.14             & 0.14             & 0.14           \\
      $\gamma_M$     &  [eV]     & 5           & 15              & 6                & 6                & 8                & 6             \\ \\
      $\Delta$E$^{Pl}$ & [eV]    & 29          & 29              & 22               & 13.2             & 13.4             & 13.4           \\
      FWHM$_L^{Pl}$  &   [eV]    & 14          & 18              & 10**             & 3.4              & 3.8              & 3.8            \\
    \hline \hline
  \end{tabular*}
	\begin{tabular*}{0.98\textwidth}{@{\extracolsep{\fill}}l}
	*Value estimated from Fermi fit of the wide scan, since broadening appears larger than for the Au film.\\
	** plus Mahan type broadening of $\alpha_M$\,=\,0.8 and $\gamma_M$\,=\,8\,eV.
	\end{tabular*}
\end{table*}

\section*{Appendix C. Band Structure}
The electronic structure calculations have been performed using the WIEN2k program package \cite{wien2k}. The generalized gradient approximation has been used for the exchange-correlation potential \cite{perdew96a}. The Ce $4f$ states have been treated as \textit{open core}. This assures the Ce ions are trivalent and a direct influence on the $5d$ density of states ($5d$-DOS) is avoided. The calculations were converged below a charge distance of 10$^{-5}$ on a 13x13x10 mesh of $k$ points with $RK_{max}=8.0$. The DOS has been calculated on a 26x26x10 mesh. Core hole effects were not taken into account \textit{in order to keep the calculation simple.}
The calculations are based on the actual 300\,K structure parameters of the respective compounds \cite{Moshopoulou_2002,Tursina2005,Muro2009_Fe,Takabatake2010}.


\end{document}